\documentclass[final]{aipproc}

\layoutstyle{6x9}

\SetInternalRegister\hbadness{8000} % pseudo latin isn't breaking very well :-)

\newcommand\doingARLO[2][]{%
  \ifx\mmref\undefined #1\else #2\fi
}

\begin{document}

\title 
      [Modelling and interpreting optical spectra of galaxies at 
	R=10\,000]
      {Modelling and interpreting optical spectra of galaxies at 
	R=10\,000}


\author{A. Lan\c{c}on}{
  address={Observatoire de Strasbourg (UMR 7550), Strasbourg, France}
}

\author{P. Ocvirk}{
  address={Observatoire de Strasbourg (UMR 7550), Strasbourg, France}
}

\author{D. Le Borgne}{
  address={Dept. of Astronomy \& Astroph., University of Toronto, Canada}
}

\author{C. Pichon}{
  address={Institut d'Astrophysique de Paris, France}
}

\author{P. Prugniel}{
  address={Observatoire de Lyon, France}
}

\author{M. Fioc}{
  address={Institut d'Astrophysique de Paris, France}
}

\author{B. Rocca-Volmerange}{
  address={Institut d'Astrophysique de Paris, France}
}

%\author{E. Thi\'ebaut}{
%  address={Observatoire de Lyon, France}
%}

\author{C. Soubiran}{
  address={Observatoire de Bordeaux, France}
}

% \copyrightholder{}
\copyrightyear  {2004}

\begin{abstract}
One way to extract more information 
from the integrated light of galaxies is to improve the spectral
resolution at which observations and analysis are carried out. 
The population synthesis code currently providing the highest 
spectral resolution is {\sc P\'egase-HR}, which was 
made available by D. Le Borgne et al. in 2004. Based on an empirical stellar
library, it provides synthetic spectra between 4000 and 6800\,\AA\ at
$\lambda/\delta\lambda=10\,000$ for any star formation history,
with or without chemical evolution. Such a resolution is particularly
useful for the study of low mass galaxies, massive star clusters,
or other galaxy regions with low internal velocity dispersions. 

After a summary of the main features of {\sc P\'egase-HR} and 
comparisons with other population synthesis codes, this
paper focuses on the inversion of optical galaxy spectra. We
explore the limits of what information
can or can not be recovered, based
on theoretical principles and extensive simulations.
First applications to extragalactic objects are shown.
\end{abstract}

\date{\today}

\maketitle

\section{Introduction}

Models for the formation of galaxies and of their internal structures
become more and more complex, triggering rapid progress in 
several related fields of astronomy. Observations are improving
(galaxy samples, spectral coverage, spatial and spectral resolution), 
population synthesis models are being
developed further, and inversion methods are being investigated with
the aim of optimizing the extraction of information from the data.  

In the field of population synthesis, one direction of progress
goes towards higher spectral resolutions. Resolving powers of the 
order of R=$\lambda/\delta\lambda$=2000 are now being used very widely
for galaxy observations. As already demonstrated with low resolution
spectrophotometric indices, such spectra allow us to break the 
age-metallicity degeneracy in the case of single stellar populations.
R=2000 is also adequate for the study of the kinematics of massive
galaxies, for which internal velocity dispersions of order 100\,km.s$^{-1}$
smooth out intrinsically finer lines. 
%Population synthesis 
%codes available for the analysis of this type of data include
%those of Vazdekis (1999) and Bruzual \& Charlot (2003).

Even higher spectral resolutions are needed to fully exploit
the information present in the light of low mass galaxies, or 
of regions of galaxies with small internal velocity dispersions
such as nuclear star clusters, low mass bulges, galaxy disks or bars. 
In the spectra of 
such objects, a wealth of strong and weak lines give access, at least in 
principle, to very detailed information about the abundances 
and kinematics of the brightest components of the stellar populations.

This paper summarizes fundamental and practical issues related to 
the synthesis and analysis of galaxy spectra at R$\simeq$10\,000
($\delta v\simeq 30$\,km.s$^{-1}$), based on {\sc P\'egase-HR}, 
the only synthesis package currently available for such a purpose
(D. Le Borgne et al. [5]).
After a description of the package in Sect.\,2, we discuss intrinsic
properties of the resulting single stellar population spectra (SSPs)
and their effects on the analysis of galaxy spectra in terms of
stellar age and metallicity distributions. We illustrate the quantitative
effects of R and of the signal-to-noise ratio (S/N) on our ability
to separate and characterize sub-populations. These results are
based on regularized inversion methods that are flexible enough to
handle complex non-linear problems such as age-velocity distributions. 
More details will be published by P. Ocvirk and collaborators.

\section{P\'egase-HR}

{\sc P\'egase-HR} is an extension of the previously available
population synthesis code {\sc P\'egase.2}, developed by M. Fioc \& 
B. Rocca-Volmerange (\url{http://www.iap.fr/pegase/pegasehr/index.html}).
Both use identical inputs in terms of stellar evolution
or yields, and if requested they ensure consistent evolution of
the metallicity in the gas and stars. {\sc P\'egase.2} used the 
Basel library of stellar spectra (Lejeune et al. [6]) to produce
low resolution spectra ranging from the UV to the near-IR
(see Rocca-Volmerange [10] for further extensions). 
{\sc P\'egase-HR} adds a zoom into the optical range, with the synthesis of
galaxy spectra at R$\simeq$10\,000 between 4000 and 6800\,\AA. 
The input high resolution
stellar library results from careful interpolations between 1959 spectra of 
1503 stars observed with the spectrograph ELODIE at Observatoire de Haute 
Provence (Prugniel \& Soubiran [9]).
%Provence\footnote{\url{http://www.obs.u-bordeaux1.fr/public/astro/CSO/elodie_library.html}}. 
%Provence\footnote{\url{http://www.obs.u-bordeaux1.fr/m2a/pages_web_m2a/web_caroline/elodie_library.html}}.
These spectra were initially observed at R=\,42\,000, which
warrants an excellent wavelength calibration and, in the smoothed
spectra actually used, a well defined constant resolved element. 
The fundamental parameters
of the library stars range from 3000\,K to 60\,000\,K in Teff, from
$-0.3$ to 5.9 in log(g) and from $-3.2$ to $+1.4$ in [Fe/H]. SSP spectra
are synthesized for $-2<$[Fe/H]$<+0.4$.

The Lick index predictions of {\sc P\'egase-HR} have been compared in 
detail with those of Bruzual \& Charlot ([3]: BC03), 
Thomas et al. ([11]: TMB03), Bressan et al. ([2]: BCT96) and 
Worthey \& Ottaviani ([15]: WO97), for ages larger
than 1\,Gyr. Indices agree well in general, and
we will only highlight a few differences here (see [5]
for details). \\
-- As all empirical libraries, the ELODIE library is affected 
by the trends found in the abundance patterns of solar neighbourhood stars\,: 
metal deficient stars tend to show an enhancement in $\alpha$-element 
abundances.  Only TMB03 apply a correction, and thus obtain
somewhat smaller Mg indices for sub-solar SSPs.\\
-- There are significant differences between authors
in the line indices of old metal-rich
populations. BC03 predict lower iron indices (Fe\,5270, Fe\,5335) 
than other groups
in this regime, although they use the same evolutionary tracks as BCT96
and {\sc P\'egase-HR}. WO97 produces a larger  Mg$_{\rm b}$ index
than other authors. The main origin for these discrepancies is clearly the 
scarcity of high metallicity low temperature giants in any of the 
empirical libraries used by the groups considered here. 
An additional cause are the uncertainties 
in the fundamental parameters of these few stars. There is a clear 
need for improvement here.\\
-- There are significant differences among the Balmer line indices
of metal-poor populations ([Fe/H]$<$-0.6). Both the tracks and the 
input spectra are likely to contribute to the differences between 
{\sc P\'egase-HR}, WO97 and TMB03. The agreement between 
{\sc P\'egase-HR} and BC03, who use the same evolutionary tracks, is better, 
although at any given metallicity
a given index would be interpreted as a slightly younger age when
using BC03.\\

At R$\simeq$10\,000, the passbands of the rectangular filters that
define the Lick indices encompass many lines of a variety of elements.
For the study of objects with internal velocity dispersions below
50\,km.s$^{-1}$, one may consider indices with passbands as 
narrow as about 2\,\AA. It is then possible to isolate the age sensitive
features from the metallicity sensitive ones, and to break the 
age-metallicity degeneracy within only a very small spectral range.
A good example is found around H$\gamma$ where two new indices,
centered about 10\,\AA\ apart, provide an excellent age-metallicity 
discrimination for single stellar populations observed at S/N$\geq$50
(Le Borgne et al. [5]). 
At lower S/N, or more generally in the presence of composite populations, many
indices must be combined. It then becomes interesting to 
work with the full spectra directly.

\section{Stellar populations from high resolution optical spectra}

The problem of extracting a star formation history from 
an integrated galaxy spectrum is notoriously ill-conditioned. There
is a large amount of redundancy between the many fluxes of
a spectrum, so that the number of independent pieces of information
is smaller than it may look. On the other hand, the redundancy makes
it possible to access this information at lower S/N ratios (all S/N
ratios in the following are given per \AA). A fundamental cause
of ill-conditioning is the slow and sometimes almost linear evolution
of SSP spectra with age\,: good quality data is needed to distinguish
a particular SSP from the linear combination of an older one and a younger one.
The mean square differences between successive spectra, even on 
a logarithmic age scale, is not constant, and therefore time resolution
is a function of age. Finally, degeneracies between age, metallicity
and extinction set limits that depend on the wavelength coverage, resolution
and S/N of the data. As a result of ill-conditioning, there is a high
risk of overinterpreting a galaxy spectrum. If one attempts to recover too 
many parameters, for instance too many independent age bins for the description
of the star formation history, the solution will dominated (and not only
contaminated) by the noise component of the data\,: it will be meaningless. 

We have adapted non parametric inversion tools to the analysis of galaxy 
spectra (Ocvirk et al. [7]).
The aim of this effort is to quantify the amount of information
that can be extracted robustly from a galaxy observation,
given a collection of SSP spectra. The tools developed
are flexible. They can be used with any population synthesis predictions,
and they can be adjusted to deal with a variety of linear or non-linear
problems. Here we will focus on results obtained with the SSP
spectra from {\sc P\'egase-HR}, which in particular allows us to
discuss the role of spectral resolution at the wavelengths
of these models.

\subsection{A test case : the star formation history at known Z}

Before considering more complex situations, it is useful to examine one of
the simplest inversion problems\,: the determination of the contributions
of SSPs of various ages in a composite population of which the metallicity
is known. The model envisaged to describe the data is a linear
combination of the SSP spectra, which are normalized to a 
common mean flux and can be arranged to form the
columns of a matrix (the "kernel"). The unknowns are the 
contributions of the various ages to the light (masses can be derived
a posteriori using the mass-to-light ratios of the SSPs).

Even at high resolution, the singular values of the kernel 
span many orders of magnitude. This 
is a clear symptom of ill-conditioning, but it also illustrates the
correlation between the noise patterns of SSPs of successive ages,
which are slightly different combinations of a finite number of
stellar spectra. For uncorrelated white noise corresponding to a S/N ratio
of a few hundred (as appropriate for {\sc P\'egase-HR}), the singular values
would not drop below a predictable minimum value. The number of singular values
above this threshold corresponds to the number of independent pieces 
of information that can be recovered reliably. For {\sc P\'egase-HR}, this
number is of about 6. Therefore, one should either restrict the 
number of independent age bins to about 6, or adopt what we call a regularized
non-parametric approach, i.e. consider a larger number of age bins 
but add smoothness constraints. The non-parametric approach,
which does not require a restrictive choice of bin limits, allows for
more freedom in the shape of the solution. For example, peaks in the 
star formation history can be age-dated precisely. 
The smoothness constraints act as
damping factors on the terms that would otherwise be dominated by noise.
Simulations lead us to advocate the differential smoothing operators 
D$_2$ or D$_3$ defined in Pichon et al. [8]. The relative importance of 
the smoothing is an inversion parameter that can be set objectively
(and thus automatically) in linear inversion problems, but still 
requires tuning based on simulations in more complex situations.

If the S/N ratio of the galaxy data available are below a few hundred,
the number of reliable pieces of information that can be recovered drops
from 6 to smaller values. Stacking data to improve the S/N ratio before
inversion is advisable, but stacking solutions derived from noisy data
does not lead to a more meaningful "average solution". The result 
would approach a predictable combination of singular solution vectors of
the kernel, i.e. a "star formation history" that says something about
intrinsic properties of the SSPs rather than about the galaxy under study. 

Systematic simulations have been run to determine how well one can
separate two dominant episodes of star formation using high resolution
optical spectra. We define the time resolution as the shortest separation
between the episodes that still leads to a clear separation between the 
two.  This time resolution is found to improve with increasing S/N,
stabilizing around 0.4\,dex at S/N=200 (0.8-0.9 dex at S/N=20). 
Spectral resolution, between R=1000 and
R=10\,000, has comparatively little impact. The random errors on the 
mean age of the episodes also decrease with increasing S/N, ranging
from about 25\,\% at S/N=20 to about 5\,\% at S/N$\simeq$200
(only cases in which a clear separation was achieved are counted;
no systematic errors due, for instance, to errors in the stellar 
libraries or the evolutionary tracks are accounted for).

Another linear problem is obtained when the model adopted 
for a galaxy spectrum is a free linear combination of SSP spectra
of different ages and metallicities. Then, solutions are two-dimensional
distributions in the age-metallicity plane. The age-metallicity degeneracy
becomes a major issue here. The study of the singular values of 
the kernel of this inversion problem shows that the number of 
independent pieces of information that can be recovered is 
larger than in the case of a single known metallicity, but not twice
as large.  For a single stellar population, the 
mean age and metallicity can be recovered precisely, but as a result
of (inevitable) smoothing a broader distribution is found, that grossly
aligns along the age-metallicity degeneracy line. Recovering 
useful distributions in the case of more complex star formation histories
is in many cases impossible at S/N ratios compatible with the {\sc P\'egase-HR}
SSPs (i.e. a few hundred); and such S/N ratios are difficult to reach
in galaxy observations. It is more reasonable to consider the restricted
problem of a single-valued age-metallicity relation, which, however,
is a non-linear problem (see below).  

In the cases discussed above, different approaches may be
taken in dealing with the continuum of the spectra. If the observations
provide a reliable flux calibration and if extinction corrections
can be made (for example on the basis of emission line measurements),
the continuum can be used. Otherwise, the inversion has to allow for
some freedom in the shape of the continuum. The problem becomes
non-linear, and is considered explicitely in the next section. 
A few conclusions of such experiments 
are relevant to the current section. The constraints on the age distribution
of the stars at a given metallicity are present in the lines, and can
be found even when the continuum is unreliable. The mean age and metallicity
determination of a single burst population can also be recovered
without exploiting the continuum 
(as expected from the literature on Lick indices).
It seems however that the direction of the age-metallicity degeneracy
depends somewhat on whether or not the continuum is considered
a relevant constraint. If it is, the metallicity sensitivity parameter
dlog(age)/dlog(Z) (defined via SSPs as an analog of the sensitivity parameter 
of Worthey [14]) is found to be smaller than in the opposite case.

\subsection{More realistic, non linear problems}

The issues of metallicity and continuum correction raised above have
shown the need for non-linear analyses. 
When metallicity is allowed to depend
on age and when the continuum is given some freedom,
the model equation is :
$$ F_{\lambda} = \int_{t_o}^{t_f}\ \Lambda(t)\ B(\lambda,t,Z(t))\ 
f(\lambda,\eta)\ {\rm d}t $$
where $F_{\lambda}$ is the galaxy spectrum,
$B(\lambda,t,Z(t))$ is an SSP spectrum at age $t$ and metallicity $Z(t)$,
$\Lambda(t)$ is the contribution of this SSP to the 
light of the galaxy,
and $f(\lambda,\eta)$ is some smooth function of wavelength that modifies
the continuum (with one or several parameters $\eta$). 
Within this framework, we have repeated the separation experiments
described in the linear case above. We find that time resolution is
significantly worse than in the case of a known metallicity. It levels at
0.8\,dex for S/N$\geq$100, starting from values 
around 1.5\,dex at S/N$\simeq$20. However, the actual ages of the 
two star formation episodes, when these are well separated, are recovered
essentially as precisely as in the single metallicity case. The 
metallicities of the two episodes are also recovered properly. 
The random errors on these metallicities are below 0.06 dex for
S/N$>$50 (again, no systematic errors are included here\footnote{Note that
it remains to be determined how the errors depend on the slope of
the age-metallicity relation used to construct the mock data.}).
They depend on S/N and on the spectral resolution, 
and R$\geq$2500 is recommended.

\begin{figure}
  \includegraphics[clip=,angle=270,width=0.8\textwidth]{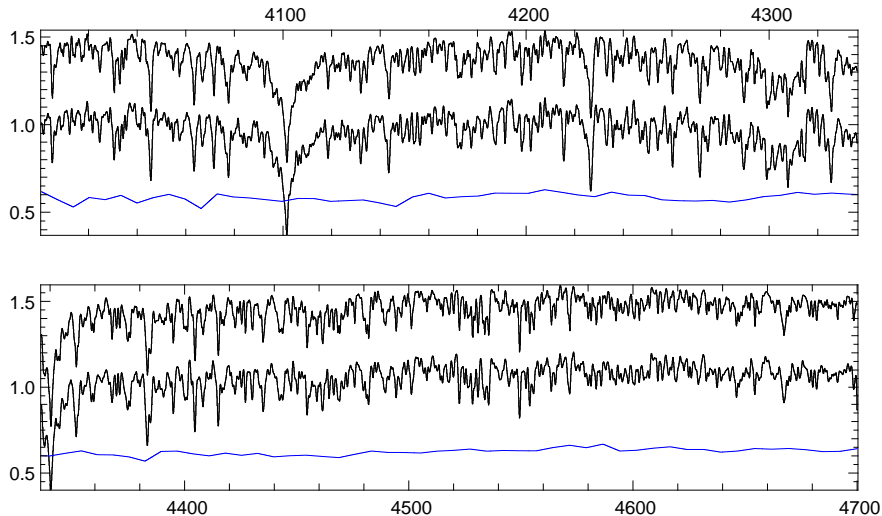}
\caption{ESO/VLT/UVES spectrum of the nuclear cluster of NGC\,300 (middle
curve) and {\sc P\'egase-HR} fit (top curve; a positive shift was
applied for clarity). To account for unknown
flux calibration errors in the data (and extinction), 
the inversion algorithm also searches for continuum
adjustments. The resulting curve shows that only very small
corrections are needed (bottom; a negative shift was
applied for clarity).}
\label{UVESfit.fig}
\end{figure}
%\begin{figure}
%  \includegraphics[width=0.49\textwidth]{NGC300-npac-SAD.ps}
%  \includegraphics[width=0.49\textwidth]{NGC300-npac-AMR.ps}
%\caption{Figure with UVES spectrum of NGC 300 central cluster}
%\end{figure}

An application to observations of the nuclear cluster of the SAd-type galaxy
NGC\,300 is shown in Fig.\,\ref{UVESfit.fig} (reduced 
data kindly provided by C.J.\,Walcher). Using wavelengths between
4000 and 4700\,\AA, we find a best fit star formation history with two 
main star formation episodes (Ocvirk et al., in preparation). 
One is younger than 1\,Gyr, the other is much older 
and has a lower metallicity. It is interesting to note that the younger 
episode is found to be dominant by Walcher et al. [12],
who used a bluer part of the spectrum (between about 3800 and 4500\,\AA)
and single-age BC03 models at solar metallicity. 
It is expected that the younger component would 
have a stronger contribution at shorter wavelengths. 
The presence of an old component would have a strong impact on the
interpretation of the dynamical mass of the cluster
(Walcher et al. [13]).
However, we stress that our results are preliminary. The metallicity 
issue as well as other sytematics need to be considered very 
carefully before final statements can be made.
The treatment of the continuum differs between
authors, and there may be differences in the detailed treatment 
and spatial coverage of the data that were used. 
P. Ocvirk is in the process of testing how accurately one can recover the star
formation histories of mock galaxies who's spectra are synthesized with
BC03 SSPs and analysed with {\sc P\'egase-HR}.
In any case, the example of NGC\,300's central cluster shows 
the quality of {\sc P\'egase-HR} fits,
and tends to support the existence of multiple star
formation episodes in some nuclear galaxy clusters, suggested 
by B\"oker et al. [1].

\begin{figure}
  \includegraphics[width=0.4\textwidth]{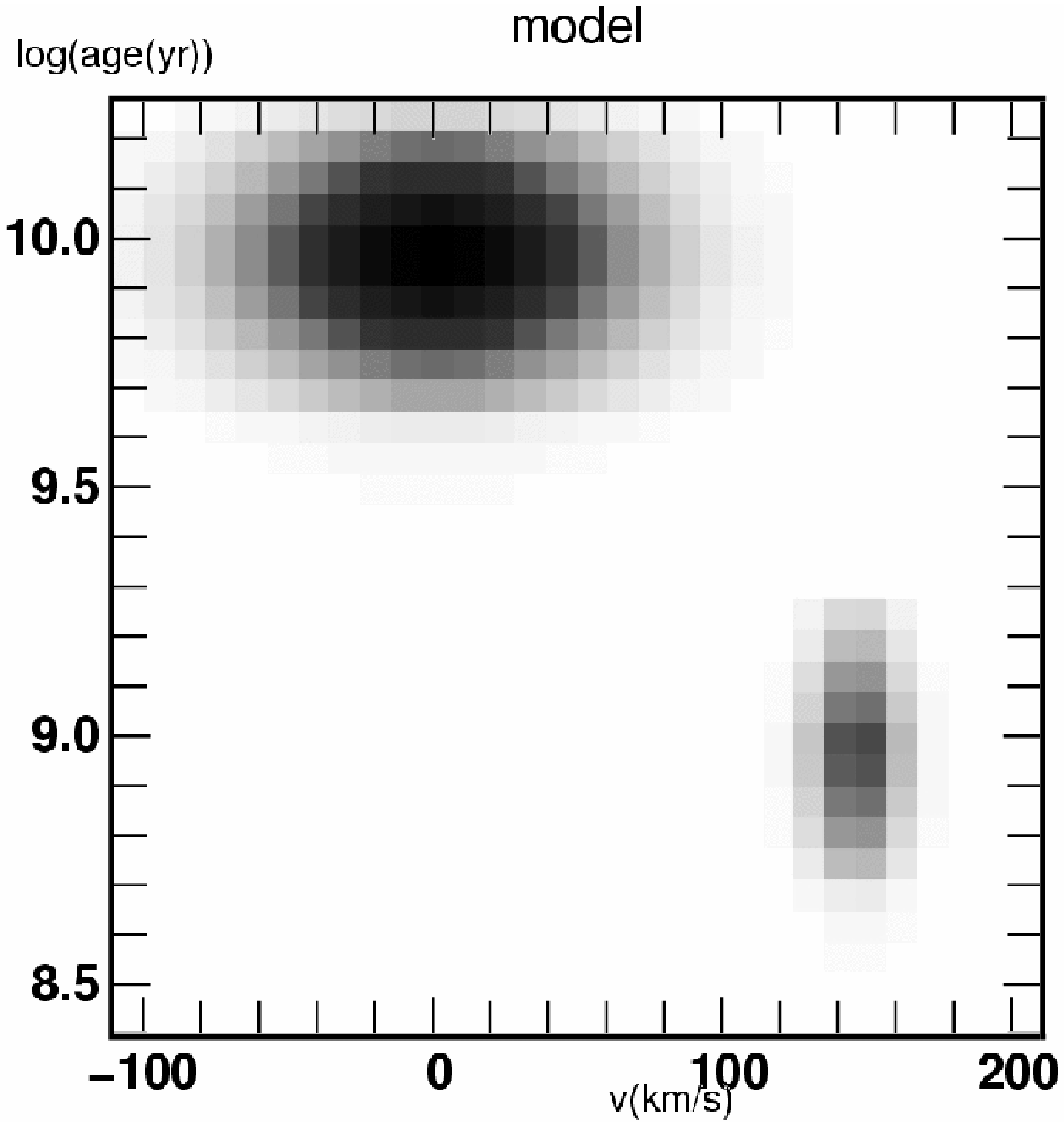}
  \includegraphics[width=0.4\textwidth]{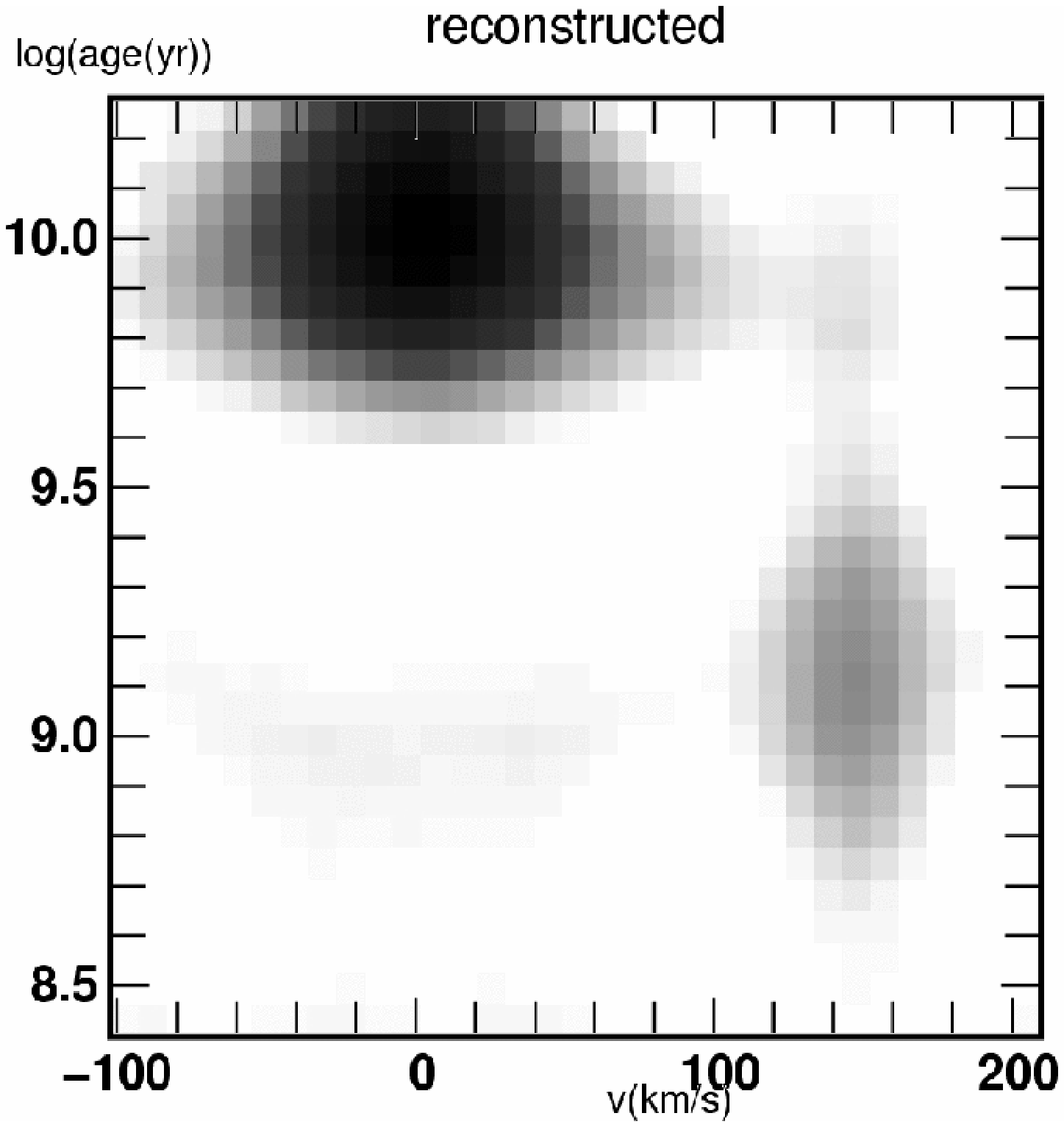}
\caption{The use of high resolution optical spectra to search for
age-velocity distributions. In this test case, the model (left)
consists of an old component with a high velocity dispersion (for example
a bulge) and a younger component at an offset velocity (for example
a disk component). The main two-dimensional structure is recovered
(right).}
\label{losvd.fig}
\end{figure}

Non-linear inversion algorithms are also necessary when the 
stellar populations of a galaxy are to be described both in 
terms of their nature (age, metallicity) and in terms of their 
kinematics.  With high resolution data of high quality, one 
may attempt to search for distributions in age and velocity
simultaneously. This may, for instance, help disentangling the 
properties of bulges and inner galaxy disks in the region of overlap.
In this framework, the model spectrum for a region of a galaxy is a 
linear combination of single stellar populations, each of which 
is allowed to have a specific velocity distribution. The solution
searched for is a two dimensional distribution in age-velocity space.
The tools developed by Ocvirk et al. [7] allow us to tackle 
this problem, with specific smoothness constraints. Figure\,\ref{losvd.fig}
illustrates preliminary results. We are in the process of determining
more systematically which configurations are favourable or prohibitive 
for this type of inversion problem. In the mean time, {\sc P\'egase-HR} 
is being used in the more traditional approach, which assumes that 
one velocity distribution adequately applies to all the stars that
contribute significantly to the light. Chilingarian et al. [4]
are analysing a sample of dwarf ellipticals in the Virgo clusters
this way, using observations with the integral field spectrograph MPFS
on the Russian 6m telescope. Their velocity maps indicate decoupled cores
or disks in a few objects. If the age or metallicity difference between
these cores and the outer regions are sufficient, age-velocity inversions
should provide interesting constraints on their formation history.

\subsection{Conclusions}

High resolution, high S/N optical spectra in good agreement with observations
of galaxies can be synthesized with {\sc P\'egase-HR} 
between 4000 and 6800\,\AA. 
While the S/N ratio (per \AA) of the data is the 
main factor determining what resolution in stellar ages can be achieved, 
the spectral resolution is critical to improve the age-metallicity
separation at these wavelengths and to study stellar kinematics. 
Regularized non-parametric inversion tools have been developed 
to analyse these (or other) data.
When the problem is set properly and the data have 
a high S/N, the random errors of the solutions are often smaller than the
systematic differences expected from the use of different population
synthesis codes. Future progress should include an improvement of the
stellar spectral libraries in certain parts of the Hertzsprung-Russell
diagram (cool giants, hot stars), 
and the simultaneous analysis of high resolution spectra and
lower resolution panchromatic energy distributions.

\begin{theacknowledgments}
We are grateful to J. Walcher for making reduced UVES spectra available to us.
We thank the organizers for a very useful and well organized meeting.
\end{theacknowledgments}

% choose bibtex style depending on layout style and options used in
% sample:

\end{document}